\documentclass{aa}  
\usepackage{graphicx}
\usepackage[varg]{txfonts}
\begin{document} 

\newcommand{\be}{\begin{equation}}
\newcommand{\ee}{\end{equation}}

\title{Constraining late-time transitions in the dark energy equation of state}

\author{C.J.A.P. Martins\inst{1,2}
\and
M. Prat Colomer\inst{3}}
\institute{Centro de Astrof\'{\i}sica, Universidade do Porto, Rua das Estrelas, 4150-762 Porto, Portugal\\
\email{Carlos.Martins@astro.up.pt}
\and
Instituto de Astrof\'{\i}sica e Ci\^encias do Espa\c co, CAUP, Rua das Estrelas, 4150-762 Porto, Portugal
\and
Institut Jaume Vicens Vives, C/ Isabel la Cat\'olica 17, 17004 Girona, Spain\\
\email{mariaprat21@gmail.com}}
\date{Submitted \today}

\abstract
{One of the most compelling goals of observational cosmology is the characterisation of the properties of the dark energy component thought to be responsible for the recent acceleration of the universe,  including its possible dynamics. In this work we study phenomenological but physically motivated classes of models in which the dark energy equation of state can undergo a rapid transition at low redshifts, perhaps associated with the onset of the acceleration phase itself. Through a standard statistical analysis we have used low-redshift cosmological data, coming from Type Ia supernova and Hubble parameter measurements, to set constraints on the steepness of these possible transitions as well as on the present-day values of the dark energy equation of state and in the asymptotic past in these models. We have also studied the way in which these constraints depend on the specific parametrisation being used. Our results confirm that such late-time transitions are strongly constrained. If one demands a matter-like pre-transition behaviour, then the transition is constrained to occur at high redshifts (effectively in the matter era), while if the pre-transition equation of state is a free parameter then it is constrained to be close to that of a cosmological constant. In any case, the value of dark energy equation of state near the present day must also be very similar to that of a cosmological constant. The overall conclusion is that any significant deviations from this behaviour can only occur in the deep matter era, so there is no evidence for a transition associated with the onset of acceleration. Observational tools capable of probing the dynamics of the universe in the deep matter era are therefore particularly important.}

\keywords{Cosmology: theory -- Dark energy -- Methods: statistical}

\titlerunning{Constraining late-time transitions in the dark energy equation of state}
\authorrunning{Martins \& Prat Colomer}
\maketitle


\section{Introduction}

Evidence for the recent acceleration of the universe has been steadily accumulating over the past two decades \citep{SN1,SN2}. Nevertheless, the properties of the dark energy hypothesised to be responsible for this acceleration (not to mention its ultimate origin) remain mostly unknown, and a plethora of possible theoretical explanations and phenomenological models has been proposed \citep{Copeland:2006wr}. In particular, one would like to know whether this is due to a cosmological constant (as first introduced, in a different context, by Einstein) or to a dynamical degree of freedom such as a scalar field.

One pragmatic observational approach to the problem consists of mapping the behaviour of dark energy as a function of redshift and trying to identify any dynamics, and this is commonly done in terms of the dark energy equation of state $w(z)=p(z)/\rho(z)$, with a constant $w_\Lambda=-1$ corresponding to a cosmological constant - see, for example, the recent review by \citet{Huterer}. Indeed, it is known that the present-day value of the dark energy equation of state, $w_0$, must be very close to $-1$ \citep{Ade:2015rim}, but the constraints at higher redshifts are weaker.

It is well known that the early universe went through several phase transitions. However, in some cosmological models these phase transitions can also occur in the more recent universe. Examples include models involving topological defects (an early suggestion being \citet{Hill:1988vm}), the vacuum metamorphosis scenario of \citet{Parker1,Parker2}, in which non-perturbative quantum effects can have observationally significant consequences at late times, and scalar field models with a non-canonical kinetic term \citep{Mortonson}. Such phase transitions would affect the properties of the universe’s contents, and in particular they could change the behaviour of the dark energy equation of state.

Several phenomenological parametrisations of the dark energy equation of state have been proposed and studied, which allow for a fast transition in the dark energy equation of state, including \citet{Bassett}, \citet{Corasaniti}, and \citet{Linder:2005ne}. Our goal in the present work is to revisit these and other more recent works \citep{Lazkoz,Jaber1,Marcondes,Jaber2}, focusing on low redshift constraints on these models, by using more recent data but also by using some simplifying physical arguments and hypotheses, to be further described in the next section.

Most of the previous analyses use a combination of low redshift data (such as Type Ia supernovas) and cosmic microwave background (henceforth CMB) measurements; it is well known that the latter are particularly constraining. In the present work we have restricted ourselves to using low redshift data: specifically we used data from Type Ia supernovas \citep{Suzuki} and measurements of the Hubble parameter \citep{Farooq}. While this leads to somewhat weaker constraints, our goal is to complement the existing literature by quantifying the constraining power of the current low redshift data. That said, we also discuss the impact of the CMB and compare our results to those of previous works.

We note that recently \citet{DiValentino} have studied a different vacuum phase transition scenario, where the model parameters are chosen such that the universe contains matter plus a cosmological constant at early times and the phase transition leads to phantom dark energy today (that is, to $w_0<-1$), which asymptotes to a cosmological constant in the far future. Relying mainly on CMB data, these authors suggest that such a model may alleviate the aforementioned Hubble constant tension. By contrast, the models we study always have canonical equations of state ($w(z)\ge-1$), so in this sense the present work complements theirs although the motivations of both are somewhat different.

\begin{figure*}
\centering
\includegraphics[width=8cm]{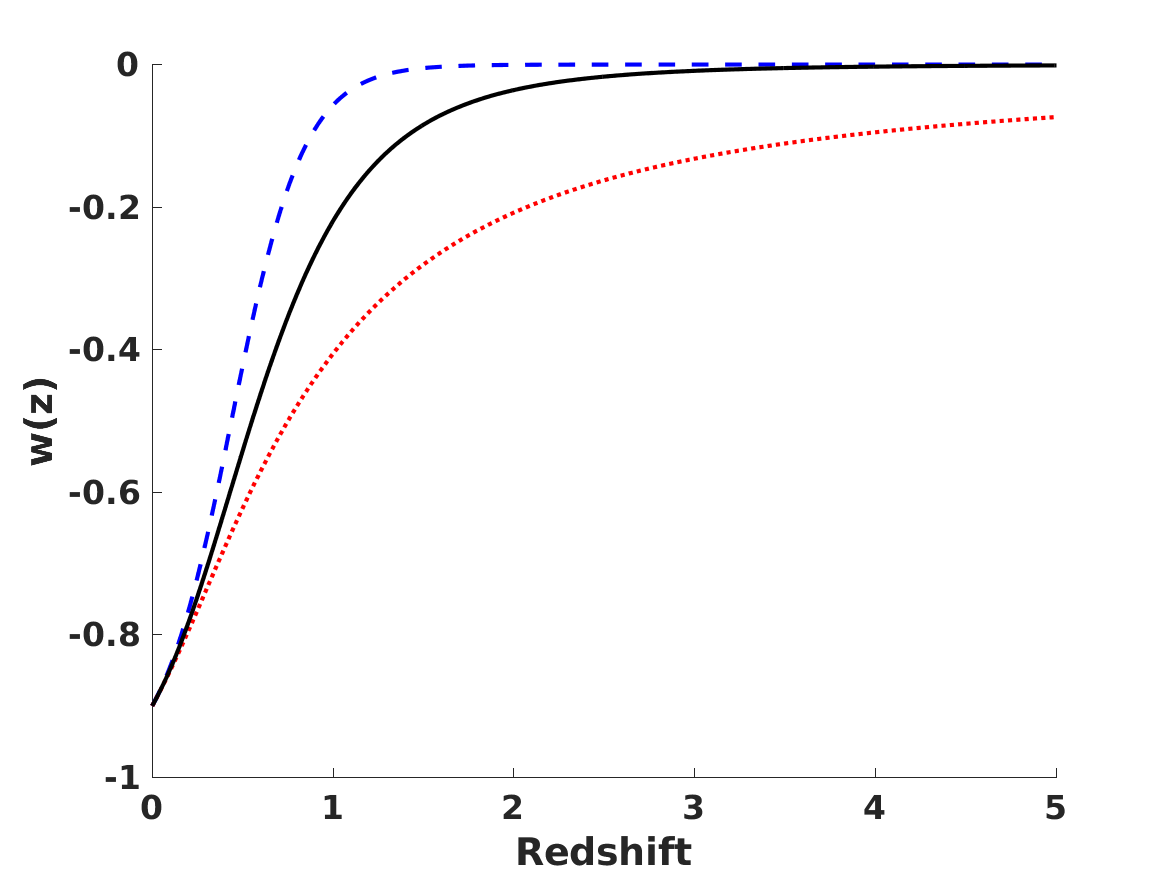}
\includegraphics[width=8cm]{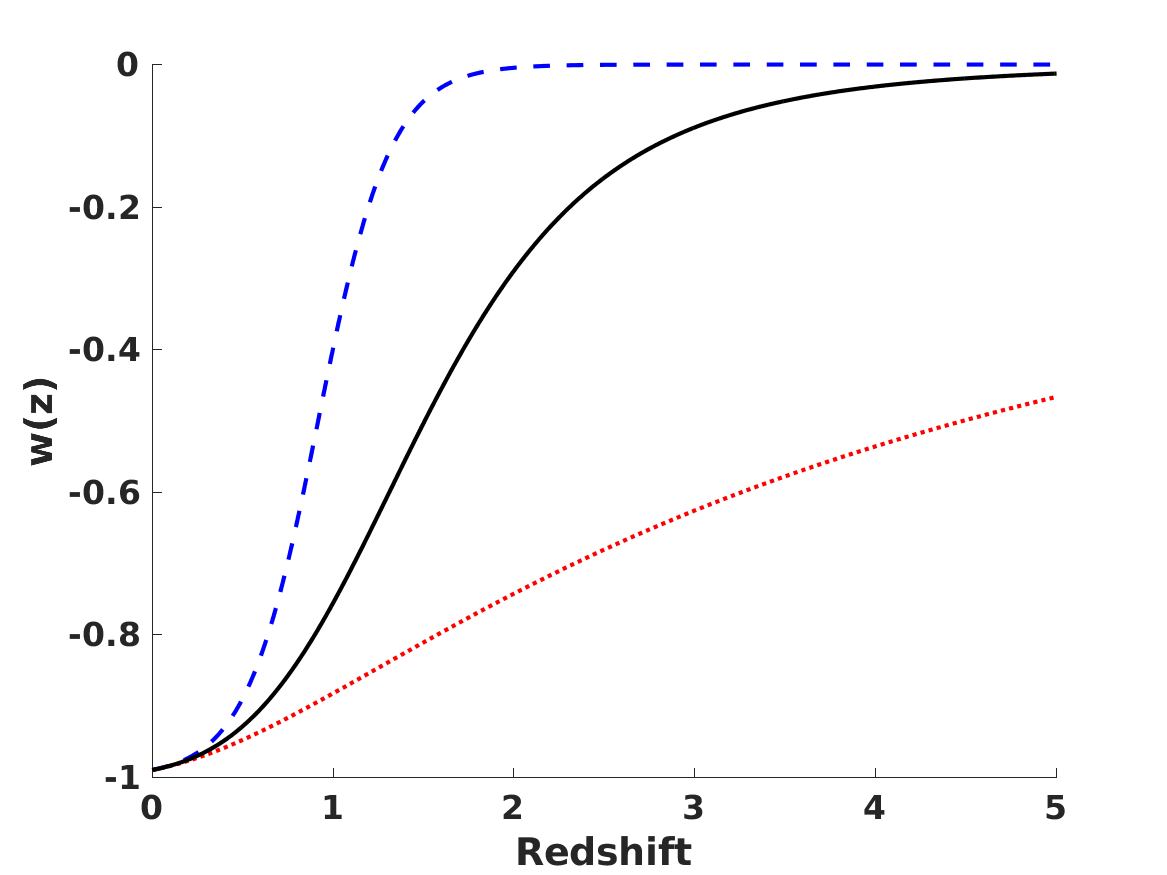}
\includegraphics[width=8cm]{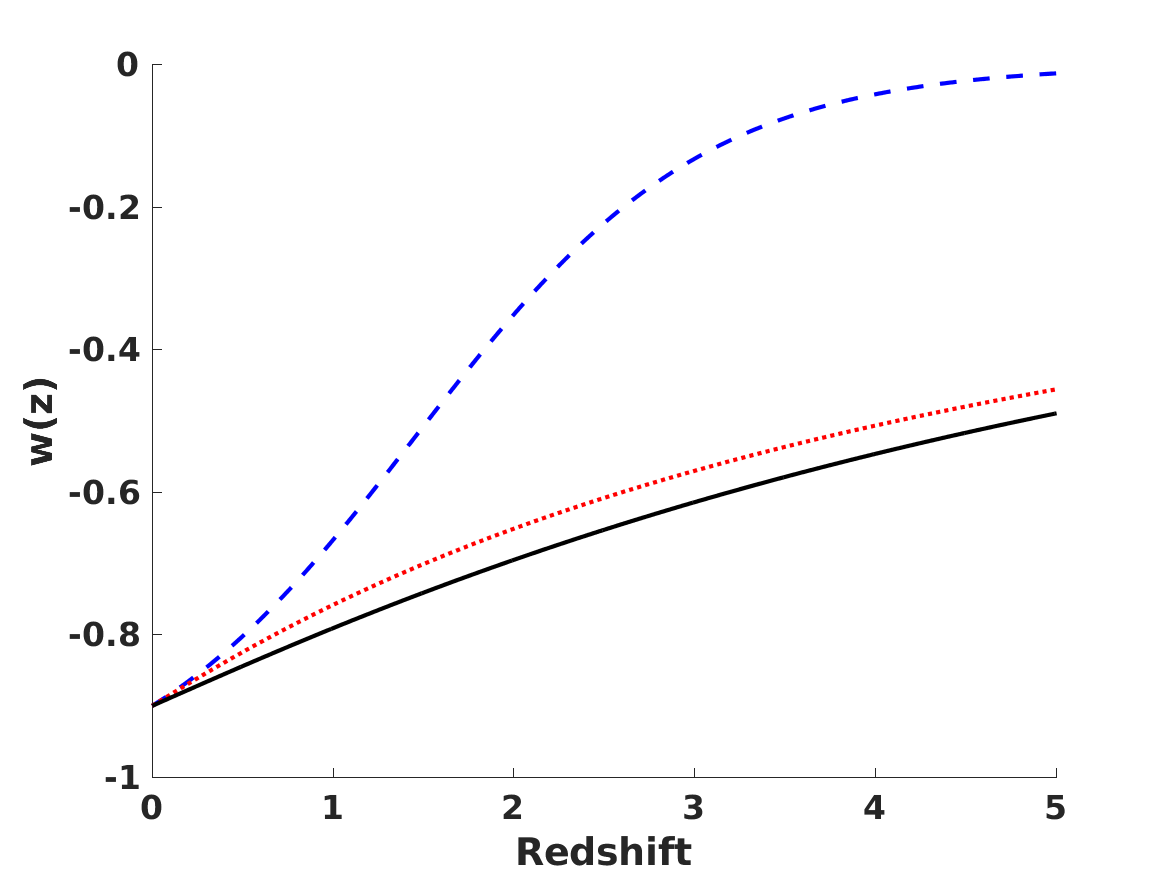}
\includegraphics[width=8cm]{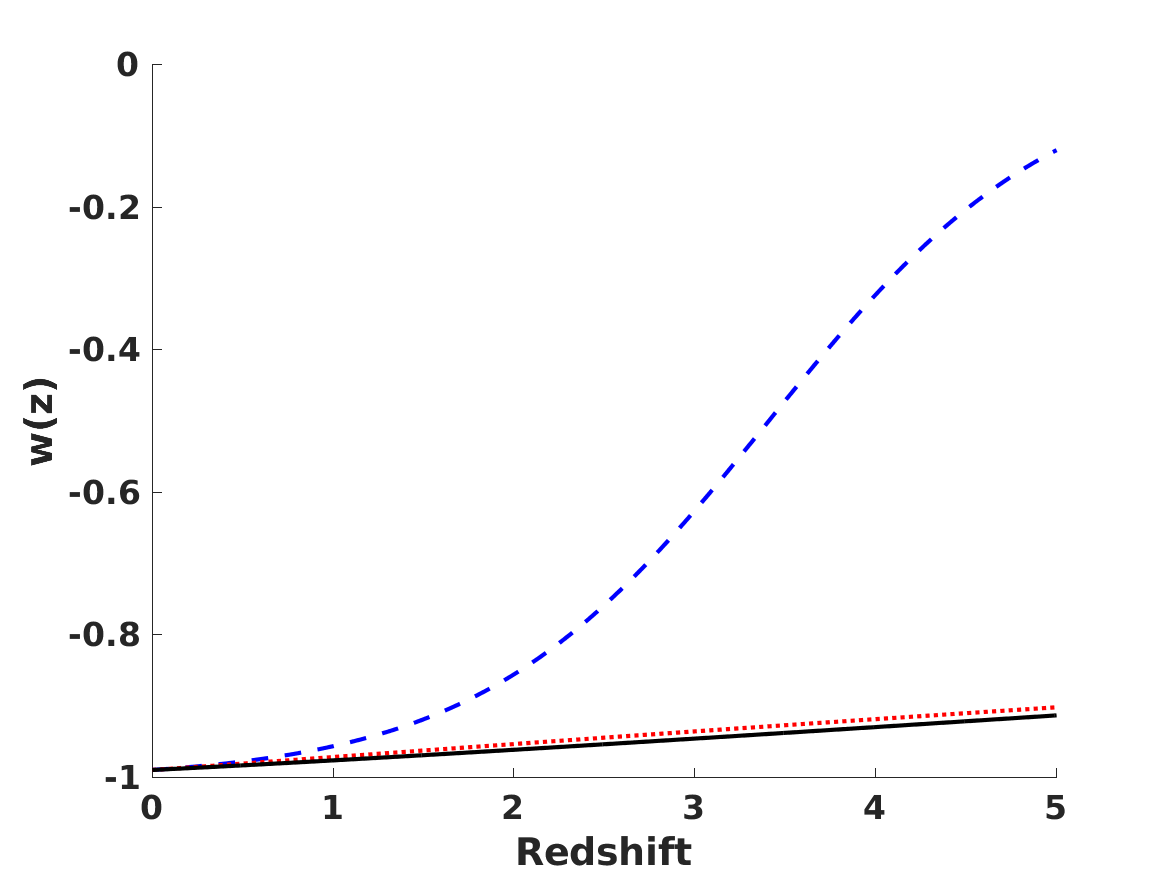}
\caption{Examples of the redshift dependence of the dark energy equation of state in the three models considered in this work, for various choices of the free parameters in the models. Blue dashed, red dotted, and black solid lines correspond, respectively, to the Models B, C and L described in the text - cf. respectively, Eqs. \ref{Modelb1}, \ref{Modelc1}, and \ref{Modell1}. The top left panel has $w_0=-0.9$ and $\Delta=0.2$, the top right one has $w_0=-0.99$ and $\Delta=0.2$, the bottom left one has $w_0=-0.9$ and $\Delta=0.8$, and the bottom right one has $w_0=-0.99$ and $\Delta=0.8$. }
\label{Fig1}%
\end{figure*}

\section{Two-parameter dark energy equation of state parametrisations}

In order to parametrise a transition in the dark energy equation of state, one generically needs four free parameters: the values of the equation of state itself at early and late times (respectively denoted $w_i$ and $w_f$), plus the characteristic redshift at which the transition occurs, $z_t$, and a transition width $\Delta$ (describing whether the transition is fast or slow). We note that the present-day value of the dark energy equation of state, $w_0$, can be expressed in terms of these four parameters, or alternatively replace one of them.

Three such phenomenological parametrisations have been described and studied in previous works:
\begin{itemize}
\item \citet{Bassett} have parametrised the transition as a function of redshift,
\be
w(z)=w_i+\frac{w_b-w_i}{1+\exp{\left[\frac{z-z_t}{\Delta}\right]}}.
\ee
We note that \citet{Bassett} state that their parameter $w_b$ corresponds to the late time behaviour of their equation of state, a parameter which we denote $w_f$ (and in their paper they indeed denote it $w_f$ in the above equation). This is only approximately correct; the exact relation between the two is
\be
w_b=w_f\left(1+e^{-(1+z_t)/\Delta}\right) - w_i e^{-(1+z_t)/\Delta}\,.
\ee
\item \citet{Corasaniti} suggest an analogous parametrisation, but expressed as a function of the scale factor,
\be
w(a)=w_0+(w_i-w_0)\frac{1+e^{a_t/\Delta}}{1-e^{1/\Delta}}\frac{1-\exp{\left[\frac{1-a}{\Delta}\right]}}{1+\exp{\left[\frac{a_t-a}{\Delta}\right]}}
\ee
\item \citet{Linder:2005ne} (see also  \citet{Linder:2007wa}) later suggested an alternative parametrisation, which has the practical advantage of allowing an analytic expression for the Hubble parameter,
\be
w(a)=w_f+\frac{w_i-w_f}{1+\left[\frac{a}{a_t}\right]^{1/\Delta}}
\ee
\end{itemize}

In what follows we have used the three parametrisations, as a means to ascertain how much the constraints derived from current data depend on this choice. On the other hand, we introduce two physically reasonable simplifying assumptions. Firstly, and considering the fact that current data can only tightly constrain two dark energy parameters (see for example the discussion in \citet{Linder:2005ne}) in the next section we restrict ourselves to two-parameter models whose dark energy equation of state behaves as matter in the asymptotic past, and as a cosmological constant in the asymptotic future: in other words, we choose two of the free parameters to have the values
\be
w_i=0
\ee
and
\be
w_f=-1\,.
\ee
The two remaining parameters can then be chosen to be $w_0$ and $\Delta$---in other words, the present-day value of the dark energy equation of state and a measure of how fast this value changes. The transition redshift can then be obtained from both of these. For the last model the $w_i=0$ assumption is relaxed later on, leading to a three-parameter model.

Secondly, we restrict ourselves to canonical models, having a dark energy equation of state with $w(z)\ge-1$; as previously mentioned, the case of phantom models, having $w(z)\le-1$, was recently studied by \citet{DiValentino}. Re-writing the above expressions in terms of our chosen free parameters, we obtain
\begin{itemize}
\item For the \citet{Bassett} parametrisation,
\be
w(z)=\frac{w_0(1-e^{-1/\Delta})}{(1+w_0)e^{z/\Delta}-(w_0+e^{-1/\Delta})}\,, \label{Modelb1}
\ee
with the transition redshift given by
\be
z_t=\Delta\log{\left[\frac{w_0+e^{-1/\Delta}}{-(1+w_0)}\right]}\,; \label{Modelb2}
\ee
we henceforth denote this specific parametrisation as Model B.
\item For the \citet{Corasaniti} parametrisation, henceforth denoted Model C, we have
\be
w(a)=\frac{w_0(1-e^{-a/\Delta})}{(1+w_0)e^{(1-a)/\Delta}-(w_0+e^{-a/\Delta})}\,, \label{Modelc1}
\ee
with the transition redshift
\be
a_t=\frac{1}{1+z_t}=\Delta\log{\left[\frac{1+w_0}{-w_0}\left(e^{1/\Delta}-1\right)-1\right]}\,; \label{Modelc2}
\ee
\item Finally, for the \citet{Linder:2005ne} model, henceforth denoted Model L, we have
\be
w(z)=\frac{w_0}{(1+w_0)(1+z)^{1/\Delta}-w_0}\,, \label{Modell1}
\ee
with a transition redshift
\be
z_t=\left(\frac{-w_0}{1+w_0}\right)^\Delta-1\,. \label{Modell2}
\ee
\end{itemize}

\begin{figure}
\centering
\includegraphics[width=8cm]{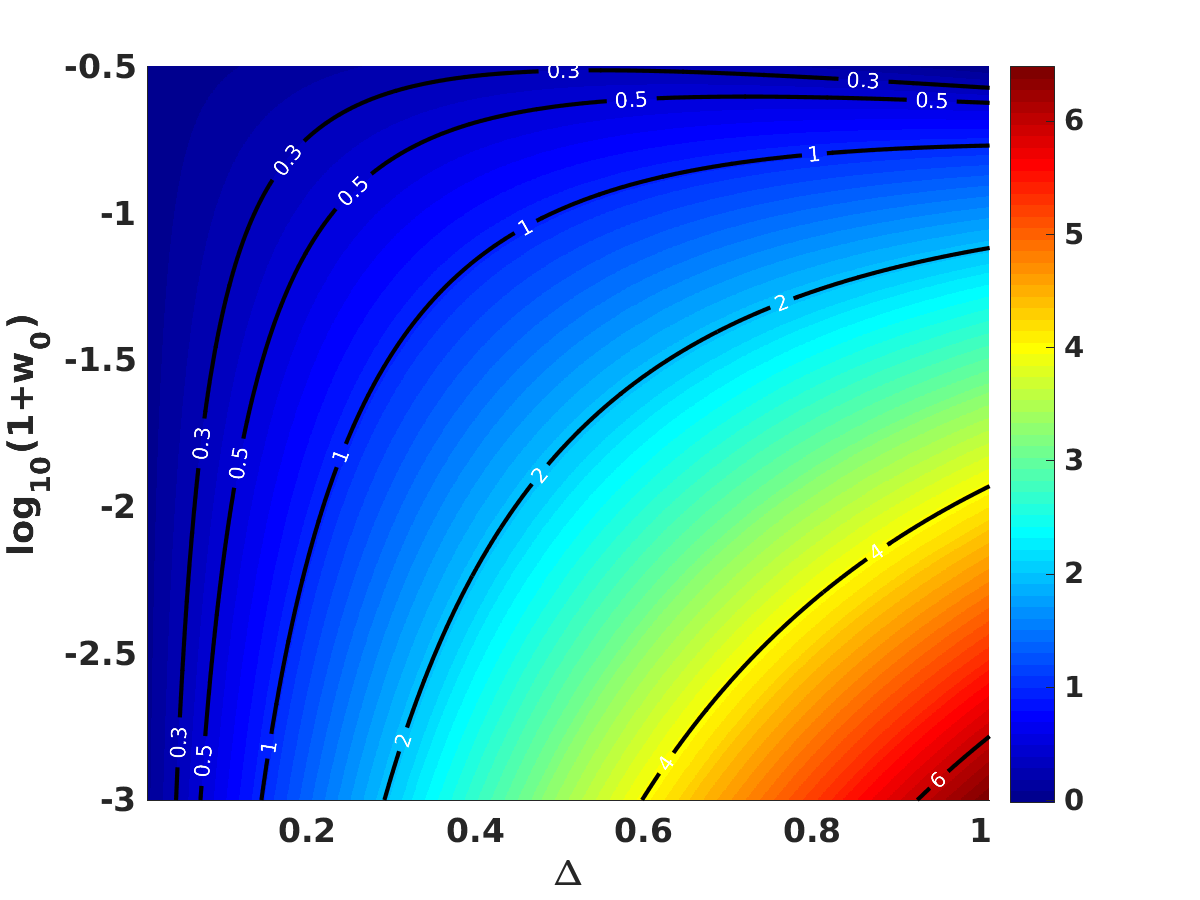}
\includegraphics[width=8cm]{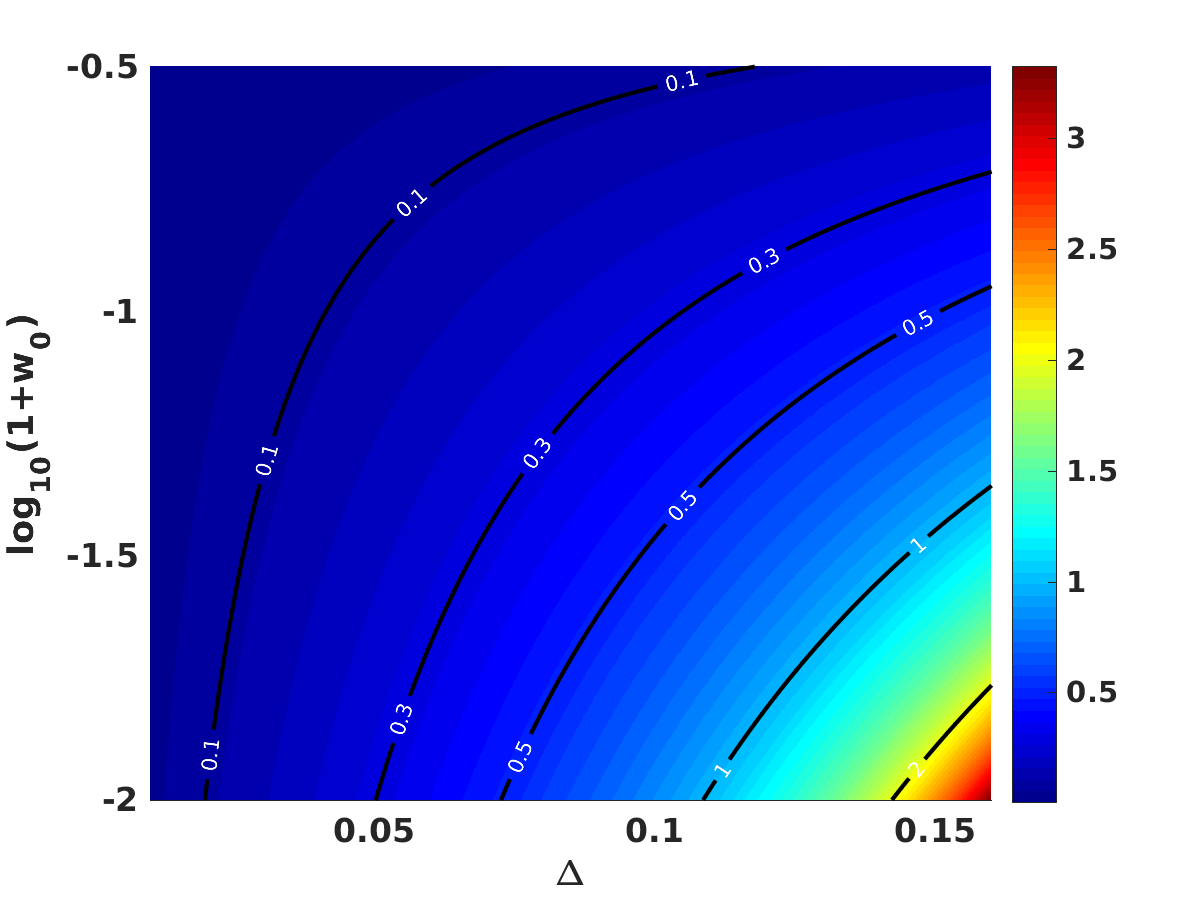}
\includegraphics[width=8cm]{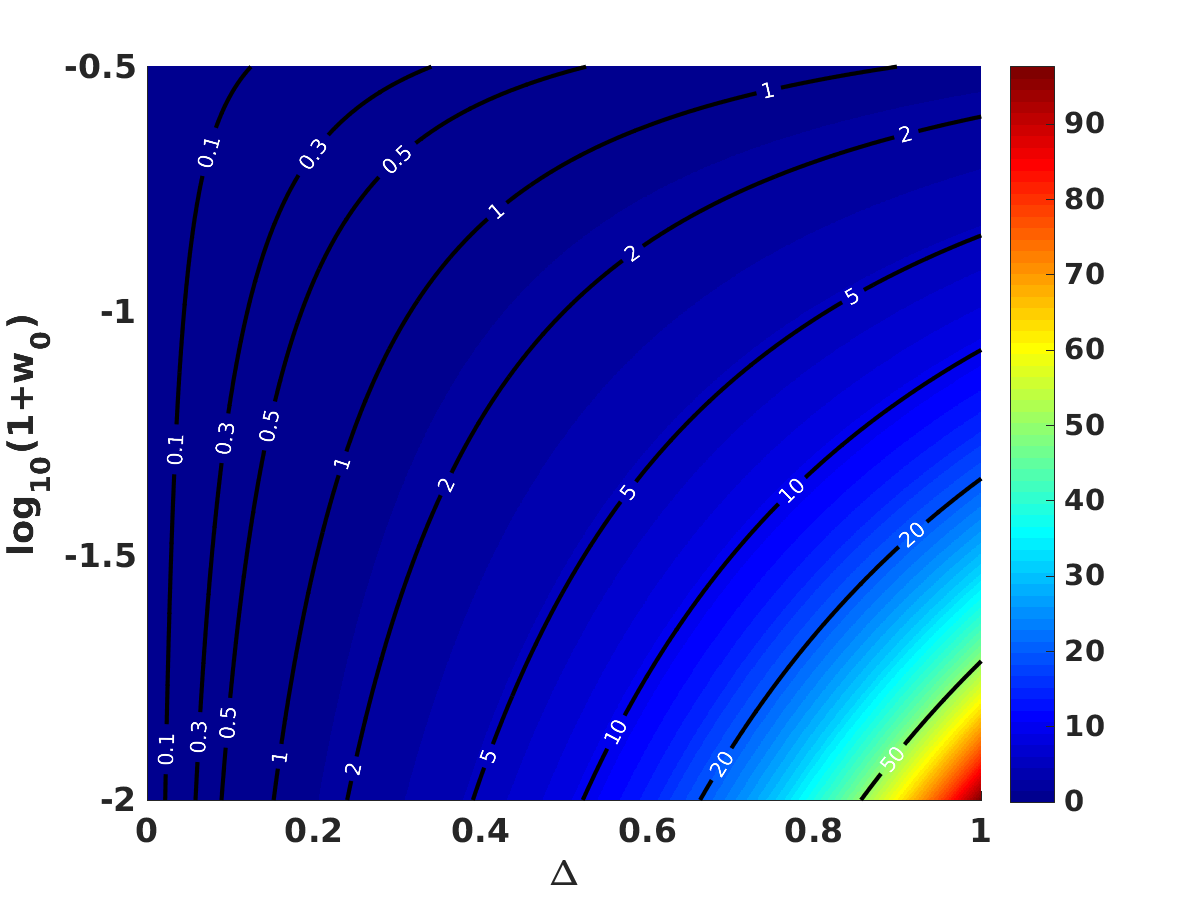}
\caption{Dependence of the transition redshift $z_t$ on the present-day value of the dark energy equation of state $w_0$ and the transition width $\Delta$ in the three models considered in this work. The top, middle and bottom panels correspond, respectively, to the Models B, C, and L described in the text - cf. respectively, Eqs \ref{Modelb1}--\ref{Modelb2}, \ref{Modelc1}--\ref{Modelc2}, and \ref{Modell1}--\ref{Modell2}. In all panels the colourmap denotes the value of $z_t$ corresponding to each pair of values of $w_0$ and $\Delta$ (we note that the colourmap and the ranges of $w_0$ and $\Delta$ are different in each panel) and some specific contours are also identified by black lines with white labels.}
\label{Fig2}%
 \end{figure}

We note that there are in principle choices of parameters for which one would obtain a negative transition redshift, implying that in such cases the effective epoch of transition only occurs in the future. However, as we see in the following section, such choices are observationally excluded. Figure \ref{Fig1} illustrates the redshift dependence of $w(z)$ in our three models, for various choices of the free parameters $w_0$ and $\Delta$, while Figure \ref{Fig2} shows how the transition redshift depends on the two free parameters; we note that different ranges of the parameters $w_0$ and $\Delta$ are depicted in the various panels of this figure, since different parameter choices will lead to reasonable values of transition redshifts in each of the models.

We emphasise that the models have in common the fact that they are two-parameter models for dynamical dark energy, described by its current value, $w_0$, and a measure of how fast this changes, $\Delta$. Naturally, $w_0$ has the same physical meaning in all three models. On the other hand, $\Delta$ plays the same qualitative role in all three models (describing how fast the dark energy equation of state evolves, near the present day), but it is clear that its quantitative meaning is different in each model, and therefore this will also be the case for the characteristic redshift of the transition, $z_t$, which can be obtained from it and $w_0$. This should also be clear from Figures \ref{Fig1} and \ref{Fig2}. Thus a direct comparison of the constraints on these parameters for the different models is instructive, but should not be taken literally.

We further assume flat Friedmann-Lema\^{\i}tre-Robertson-Walker models, so that the present-day values of the matter and dark energy densities (expressed as a function of the critical density) satisfy $\Omega_m+\Omega_{dark}=1$. We neglect the contribution of the radiation density, since we will only be concerned with low-redshift data. Thus the Friedmann equation will have the general form
\be
\frac{H^2}{H_0^2}=\Omega_m(1+z)^3+(1-\Omega_m)\exp{\left[3\int_0^z\frac{1+w(z')}{1+z'}dz'\right]}\,;
\ee
for the particular case of Model L, the substitution of Equation \ref{Modell1} leads to the analytic expression
\be
\frac{H^2}{H_0^2}=\Omega_m(1+z)^3+(1-\Omega_m)\left[(1+w_0)(1+z)^{1/\Delta}-w_0\right]^{3\Delta}\,.
\ee

\section{Low-redshift constraints on two-parameter models from current data}

Most previous works studying these scenarios constrain them using a combination of low-redshift data (such as Type Ia supernovas) and CMB measurements. It is well known that in such cases the latter will carry most of the statistical weight in the analysis, given its high accuracy and the fact that the combination of the two observables encompasses a very wide redshift range. In the present work we will restrict ourselves to using low-redshift data, quantifying how constraining this data currently is. In the discussion section we will compare our results to those of other recent works which do use the CMB.

Specifically, we use the recent Union2.1 catalogue of Type Ia supernovas \citep{Suzuki} together with the compilation of measurements of the Hubble parameter by \citet{Farooq}. The latter has the advantage of extending our redshift lever arm, since it includes measurements up to redshift $z\sim2.36$, while the supernova data is all at $z<1.5$ (and indeed mostly at $z<1$). On the other hand, we emphasise that this compilation of Hubble parameter measurements is heterogeneous: some of the measurements come from galaxy clustering and baryon acoustic oscillations (henceforth BAO) observations, while others come from the so-called cosmic chronometers or differential age method \citep{Jimenez:2001gg}. We note that it is not currently clear that possible systematics issues of the cosmic chronometers method are well understood and under control \citep{LiuLu,Vazdekis,Concas,Corredoira}.

We have carried out a standard statistical likelihood analysis, assuming a flat logarithmic prior on $(1+w_0)$, specifically $\log_{10}{(1+w_0)}=[-3,-0.5]$, and a uniform prior on the transition width, $\Delta=]0,1]$. Our choice of prior for $w_0$ is motivated by the fact that (as will be seen in what follows) the data strongly prefers a value of very close to that of a cosmological constant, and that for values smaller than $\log_{10}{(1+w_0)}=-3$ the phase space volume has almost the same likelihood. It should be noted that this choice of a logarithmic prior for a parameter whose value is near zero may affect the resulting posterior, due to the effect of prior volume weighting.

We assume a flat universe, and further fix the present-day value of the matter density to be $\Omega_m=0.3$, in agreement with both CMB and low-redshift data \citep{Ade:2015xua,Abbott:2017wau,Scolnic:2017caz,Jones:2017udy}. We initially fix the Hubble constant to be $H_0 = 70$ km/s/Mpc, but we will subsequently relax this assumption.

\begin{figure}
\centering
\includegraphics[width=8cm]{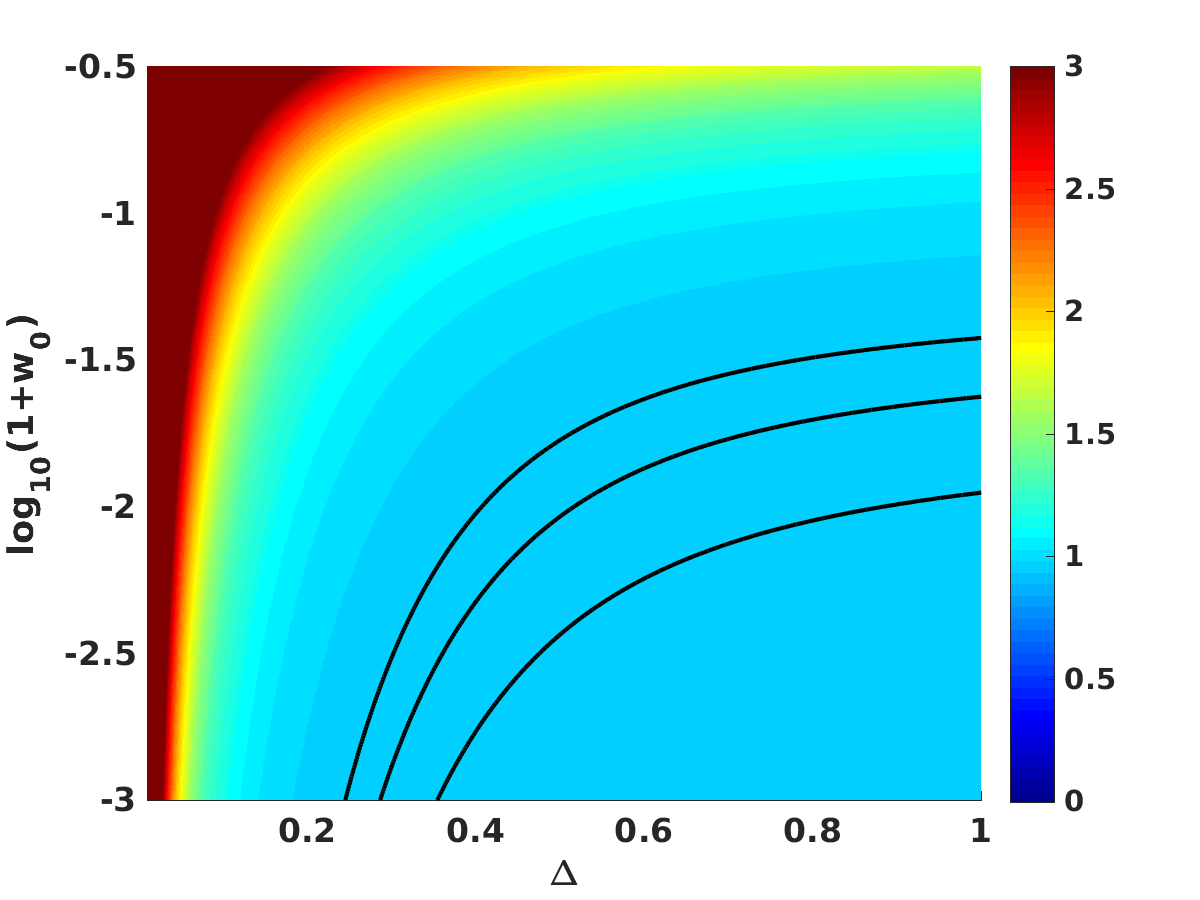}
\includegraphics[width=8cm]{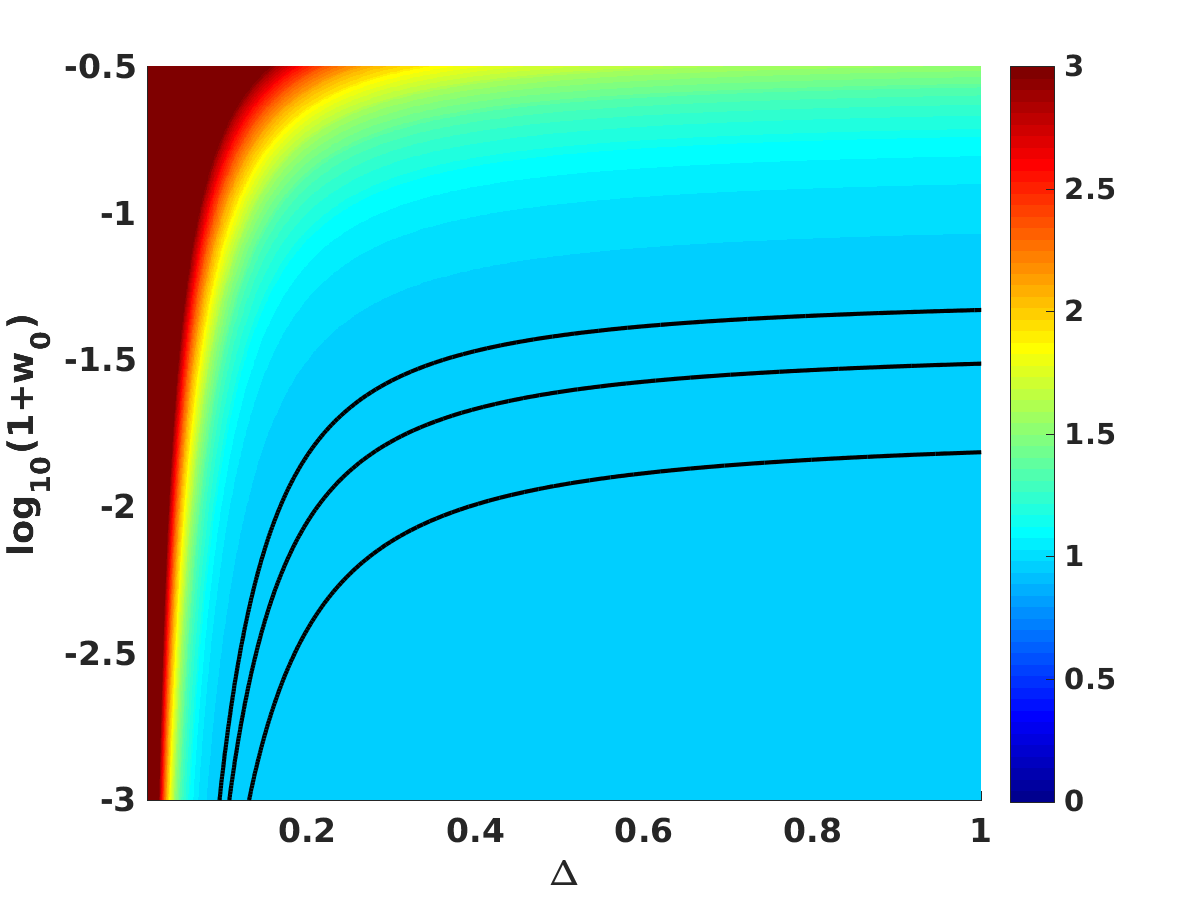}
\includegraphics[width=8cm]{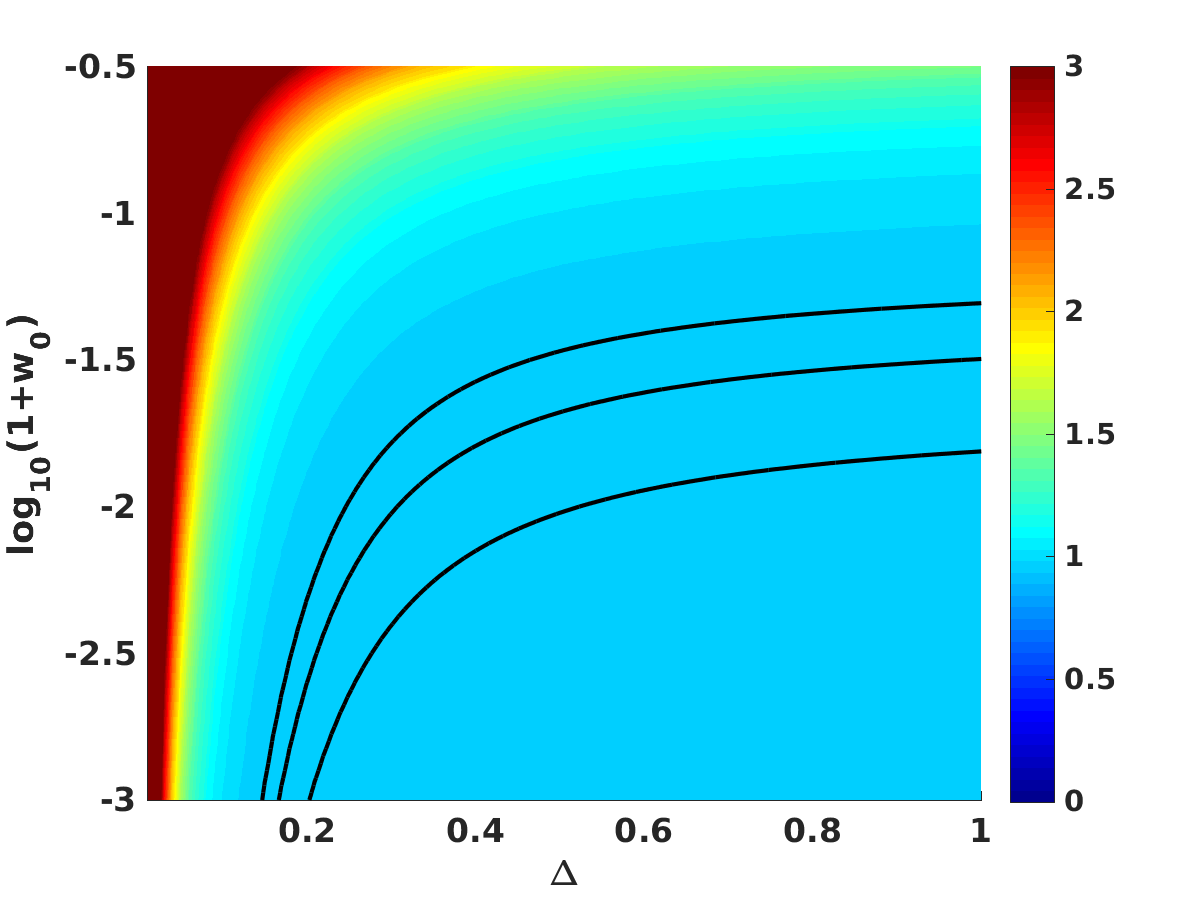}
\caption{Constraints in the ($\Delta$,$w_0$) plane for models with a late-time phase transition in the dark energy equation of state. The top, middle and bottom panels correspond, respectively, to Models B, C, and L described in the text. In all cases the Hubble constant has been kept fixed at $H_0=70$ km/s/Mpc, the black contours denote the one, two and three-sigma confidence levels, and the colourmap depicts the reduced chi-square of the fit for each set of model parameters (a dark red colour corresponds to a reduced chi-square of three or larger).}
\label{Fig3}%
\end{figure}
\begin{figure}
\centering
\includegraphics[width=8cm]{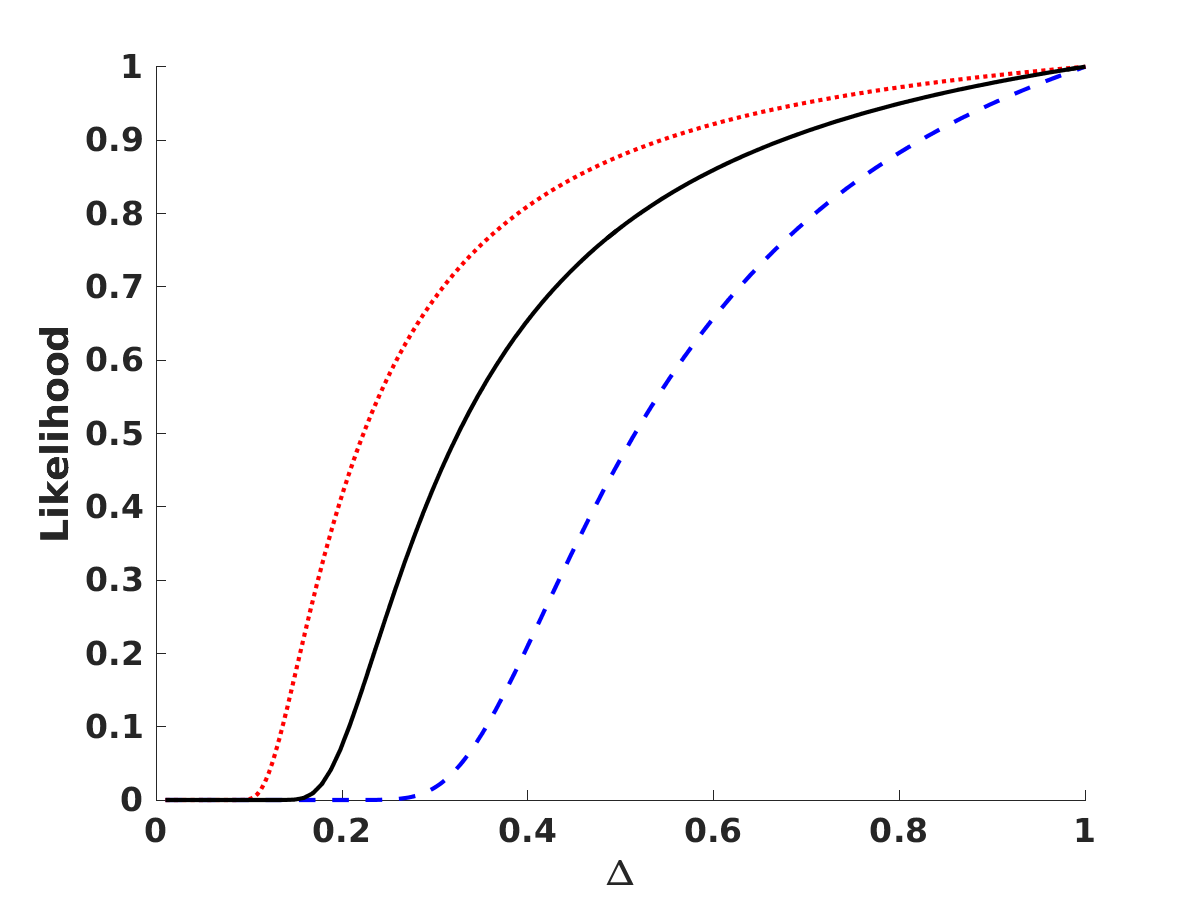}
\includegraphics[width=8cm]{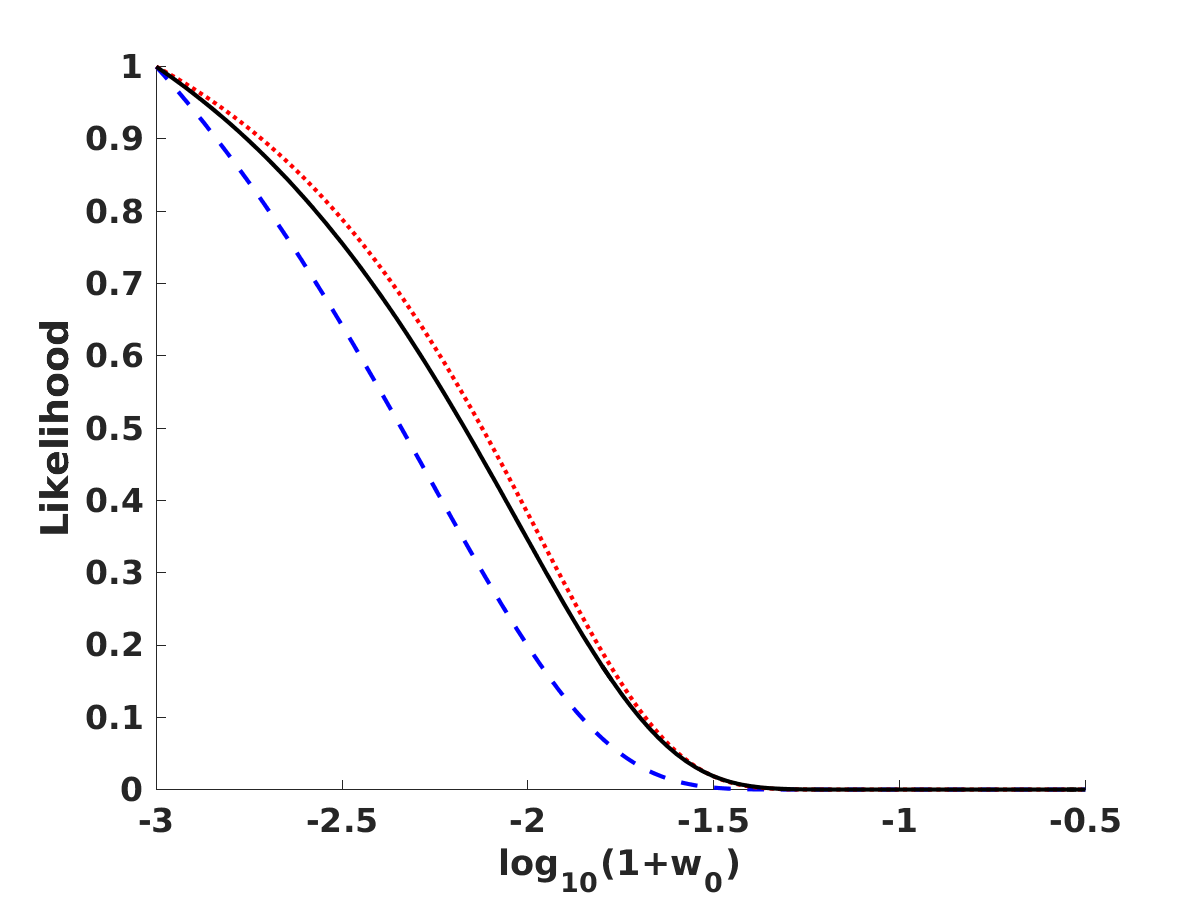}
\caption{Posterior likelihoods for the transition width $\Delta$ and the present-day dark energy equation of state $w_0$, marginalising the other, for models with a late-time phase transition in the dark energy equation of state. The blue dashed, red dotted, and black solid lines correspond, respectively, to Models B, C, and L described in the text. In all cases the Hubble constant has been kept fixed at $H_0=70$ km/s/Mpc.}
\label{Fig4}%
\end{figure}

Figure \ref{Fig3} shows the constraints on the three models in the two-dimensional ($\Delta$,$w_0$) plane, including the reduced chi-square for the comparison of the models (with each choice of parameters) with our data sets. One sees that the data strongly prefers a present equation of state close to a cosmological constant, as well as a comparatively large transition width. In other words, any large deviations from $w_\Lambda=-1$ are only allowed at large redshifts. Thus, despite the fact that we have only used relatively low redshift data, we effectively constrain such deviations to happen only well into the matter era.

The one-dimensional posterior likelihoods for $\Delta$ and $w_0$ are shown in Figure \ref{Fig4}, from which we can infer the dependence of the constraints on the choice of parametrisation for the transition. Broadly speaking, Model B generally leads to steeper transitions for a given choice of model parameters than Model C (cf. the examples shown in Figure \ref{Fig1}), and therefore the former model leads to more stringent constraints than the latter. The behaviour of Model L to some extent interpolates between the other models (for faster transitions it is closer to Model B, while for slower ones it is more similar to Model C) and therefore leads to intermediate constraints. The two-sigma ($95.4\%$ confidence level) constraints on the two parameters are summarised in Table \ref{table1}.

\begin{table}
\caption{Two-sigma ($95.4\%$ confidence level) constraints on the dark energy equation of state parameters $w_0$ and $\Delta$ (marginalised over the other), for the various two-parameter dark energy models described in the text.}
\label{table1}
\centering
\begin{tabular}{| c | c | c c |}
\hline
Model & $H_0$ & $\Delta$ & $\log_{10}(1+w_0)$  \\
\hline
Model B & Fixed & >0.37 & <-1.91 \\
Model C & Fixed & >0.14 & <-1.73 \\
Model L & Fixed & >0.22 & <-1.75 \\
\hline
Model L & Marginalised & >0.22 & <-1.73 \\
\hline
\end{tabular}
\end{table}

\begin{figure}
\centering
\includegraphics[width=8cm]{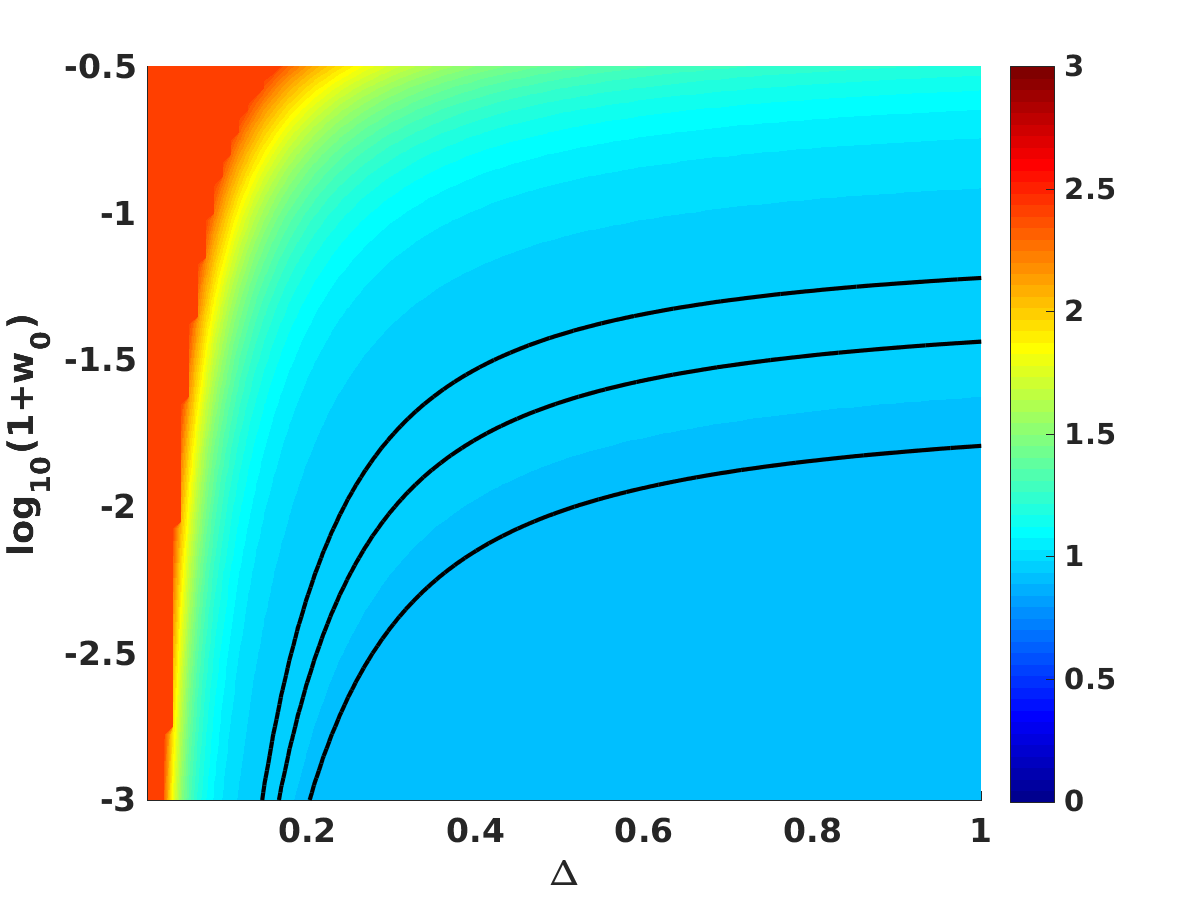}
\includegraphics[width=8cm]{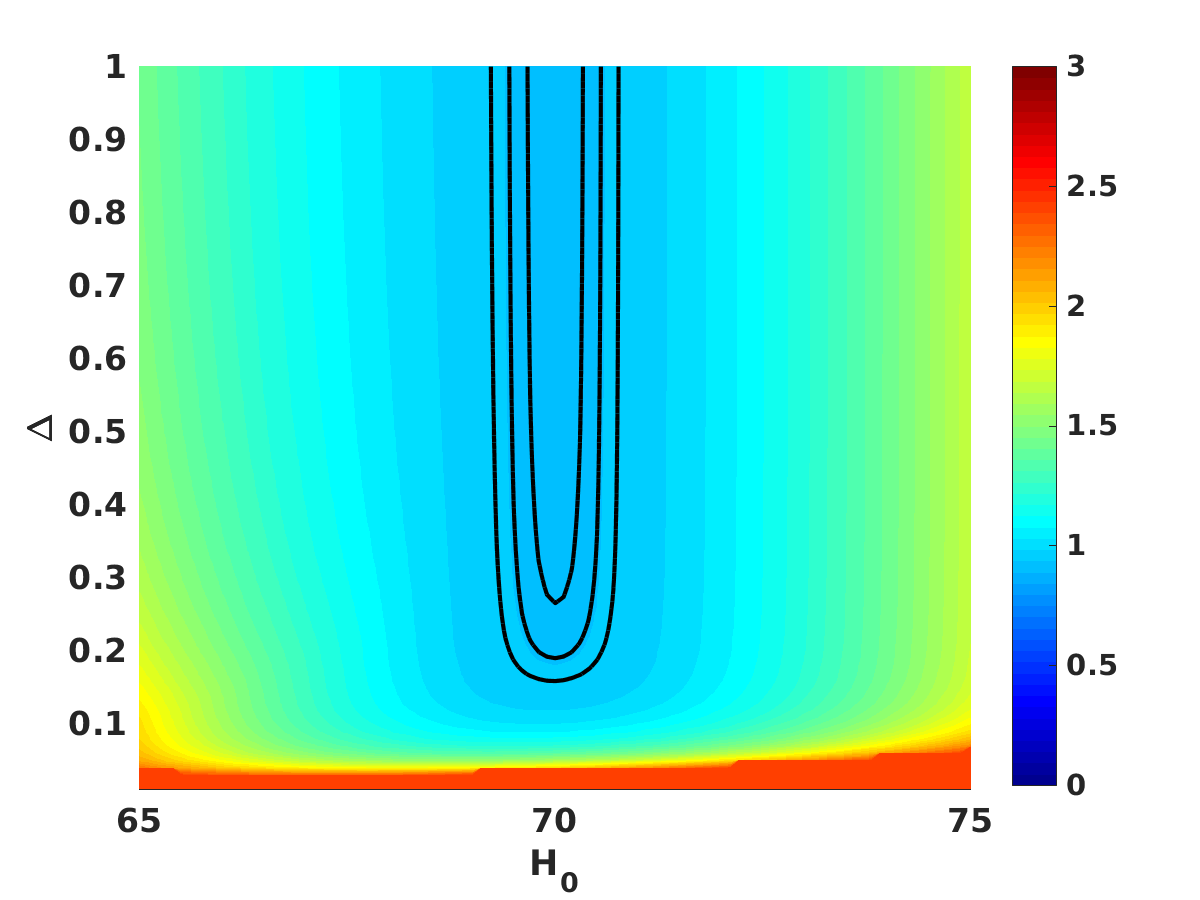}
\includegraphics[width=8cm]{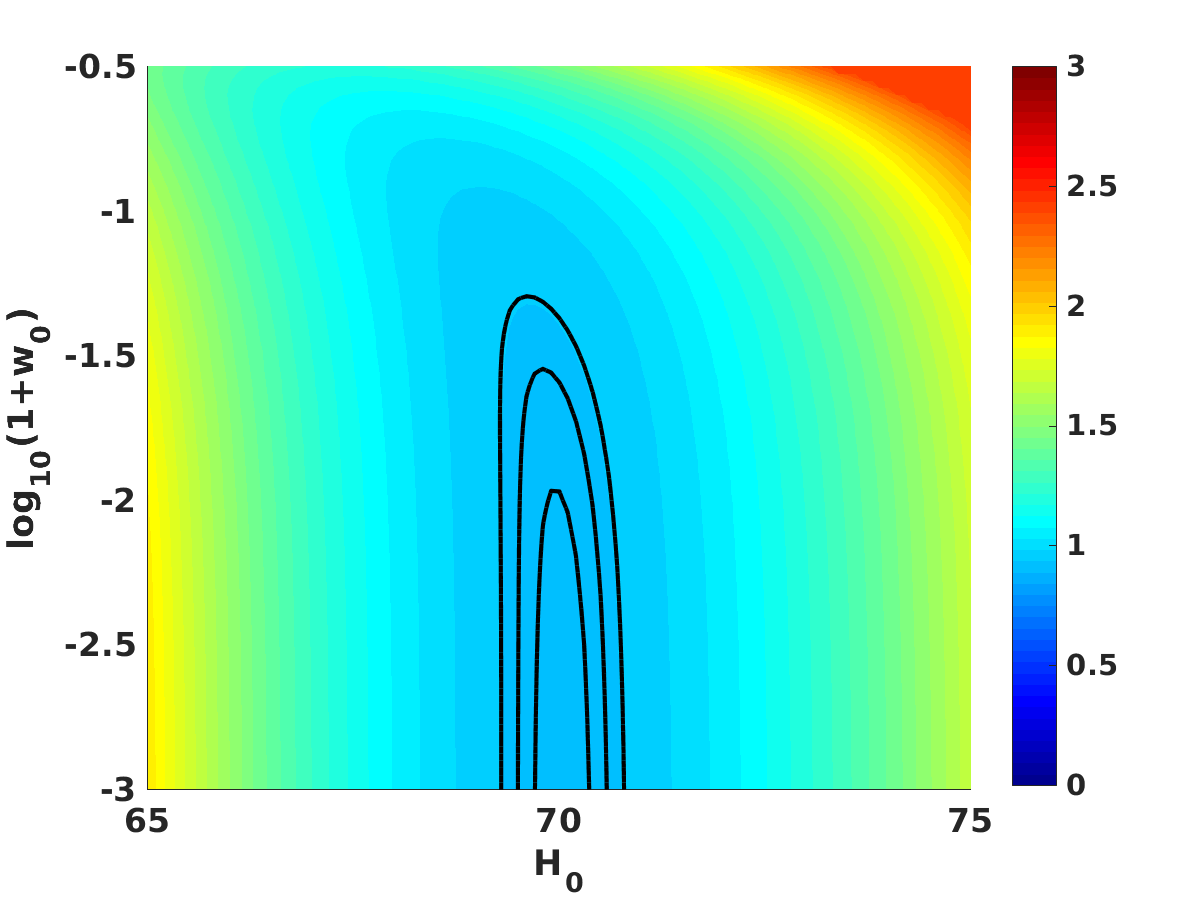}
\caption{Constraints in the ($\Delta$,$w_0$), ($H_0$, $\Delta$) and ($H_0$,$w_0$) planes (top, middle and bottom panels, respectively) for Model L, with the Hubble constant being a free parameter and the third parameter marginalised. In all cases the black contours denote the one, two and three-sigma confidence levels, while the colourmap depicts the reduced chi-square of the fit for each set of model parameters (a dark red colour corresponds to a reduced chi-square of three or larger).}
\label{Fig5}%
\end{figure}

Study of the impact of the assumption of a fixed Hubble constant is relevant to the present work. We have done this for Model L, assuming a uniform prior between  $H_0=65$ km/s/Mpc and $H_0=75$ km/s/Mpc, which span the range of reasonable values given the currently available data \citep{Ade:2015xua,Riess:2016jrr}. The results are shown in Figure \ref{Fig5}, and also shown for comparison in the last line of Table \ref{table1}. We see that this has a minimal effect on the constraints (making them weaker by a very small amount), the reason being that both data sets being used are fully compatible with the previously chosen value of $H_0=70$ km/s/Mpc.

\section{Three-parameter extension}

Our analysis in the previous section does contain one caveat. Our choice of $w_i=0$ in the above analysis implies that at early times we have an additional matter-like component\footnote{We thank the anonymous referee for emphasising this point.}. In other words, such models have a non-zero amount of early dark energy \citep{Wetterich}, which is constrained by CMB experiments such as Planck to sub-percent level \citep{Ade:2015rim}, while the constraints in the previous section would allow values as high as about ten percent. In the current section we therefore relax this assumption, and repeat the analysis for the three-parameter extension of the L model, with $w_i$ as an additional free parameter.

In this case Equation \ref{Modell1} becomes
\be
w(z)=\frac{w_i(1+w_0)(1+z)^{1/\Delta}-(w_i-w_0)}{(1+w_0)(1+z)^{1/\Delta}+(w_i-w_0)}\,; \label{Modell13}
\ee
we note that if either $w_0=-1$ or $w_i=-1$ this parametrisation reduces to $w(z)=-1$ throughout. The effective redshift is now
\be
z_t=\left(\frac{w_i-w_0}{1+w_0}\right)^\Delta-1\,, \label{Modell23}
\ee
while the Friedmann equation becomes
\be
\frac{H^2}{H_0^2}=\Omega_m(1+z)^3+(1-\Omega_m)\left[\frac{1+w_0}{1+w_i}(1+z)^{1/\Delta}+\frac{w_i-w_0}{1+w_i}\right]^{3\Delta(1+w_i)}\,.
\ee

We can now repeat the analysis of the previous section for this extended model. We still fix the cosmological parameters $\Omega_m=0.3$ and $H_0 = 70$ km/s/Mpc, and maintain our assumptions of a flat logarithmic prior on $(1+w_0)$, $\log_{10}{(1+w_0)}=[-3,-0.5]$, and a uniform prior on the transition width, $\Delta=]0,1]$. As for the early-time value of the dark energy equation of state, $w_i$ we separately consider the cases of a flat prior, $w_i=]-1,0]$, or a logarithmic prior, $\log_{10}{(1+w_i)}=[-3,0]$, again as a means of checking how sensitive the results are to these choices.

\begin{figure*}
\centering
\includegraphics[width=8cm]{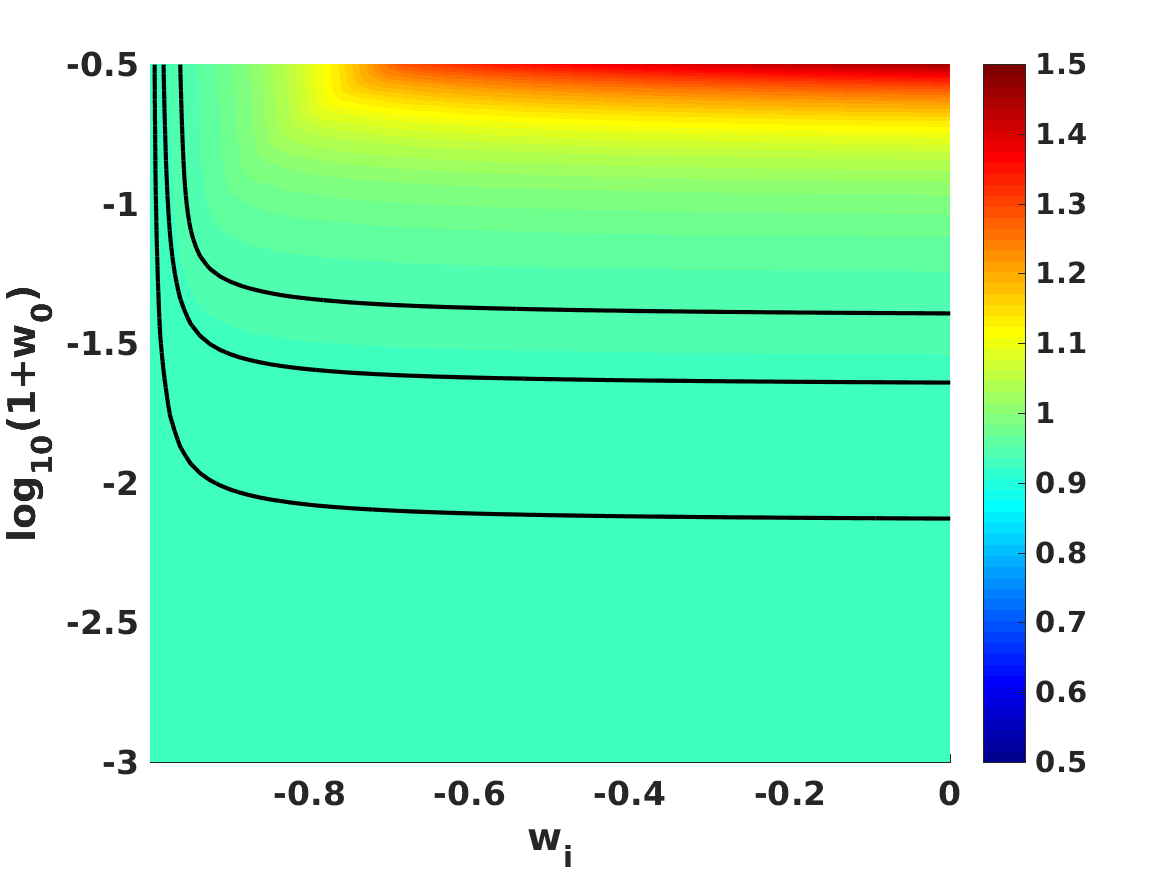}
\includegraphics[width=8cm]{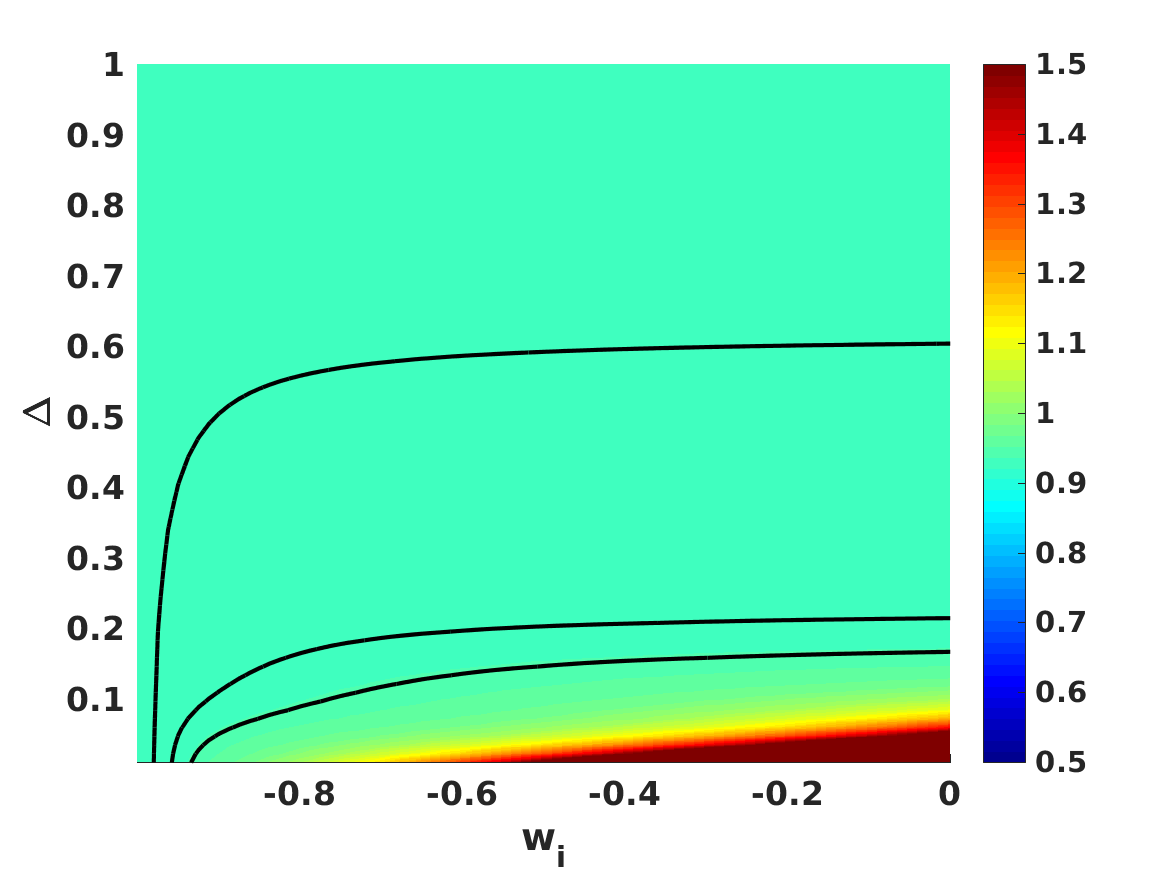}
\includegraphics[width=8cm]{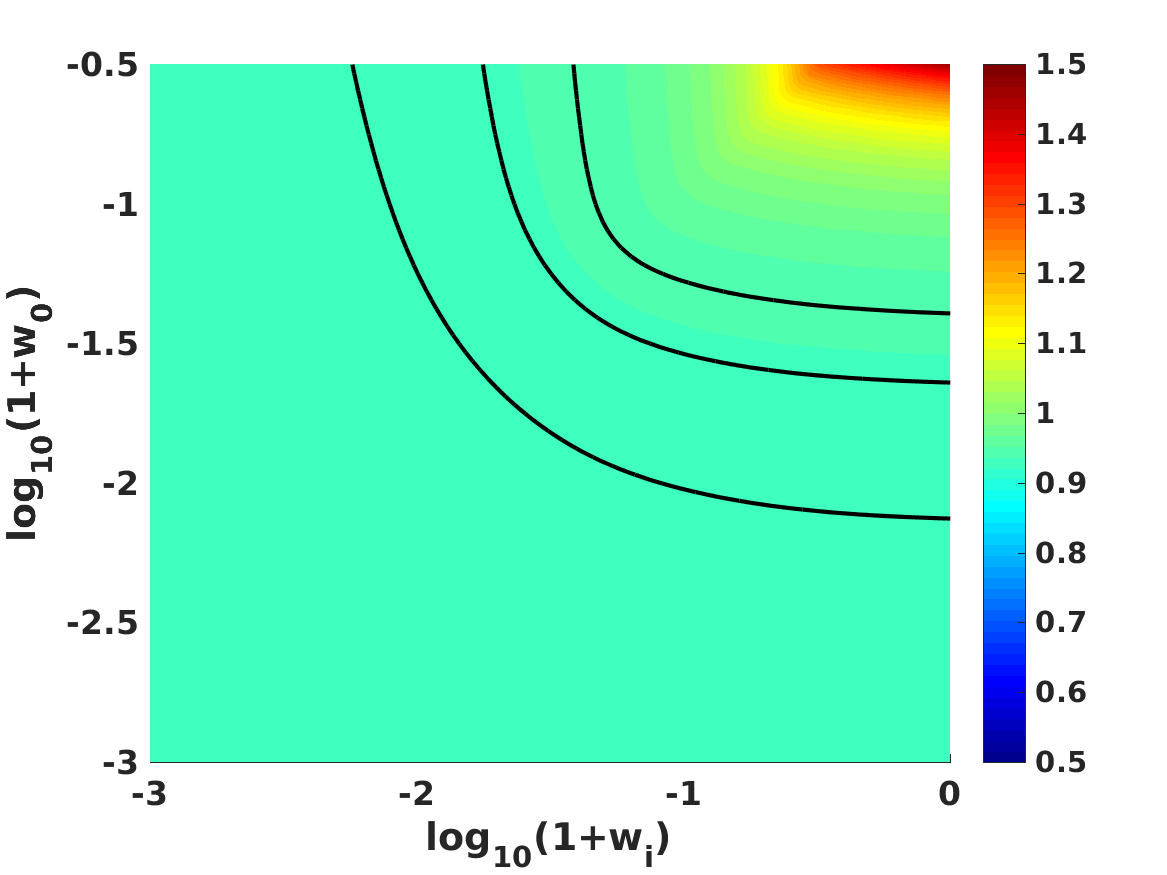}
\includegraphics[width=8cm]{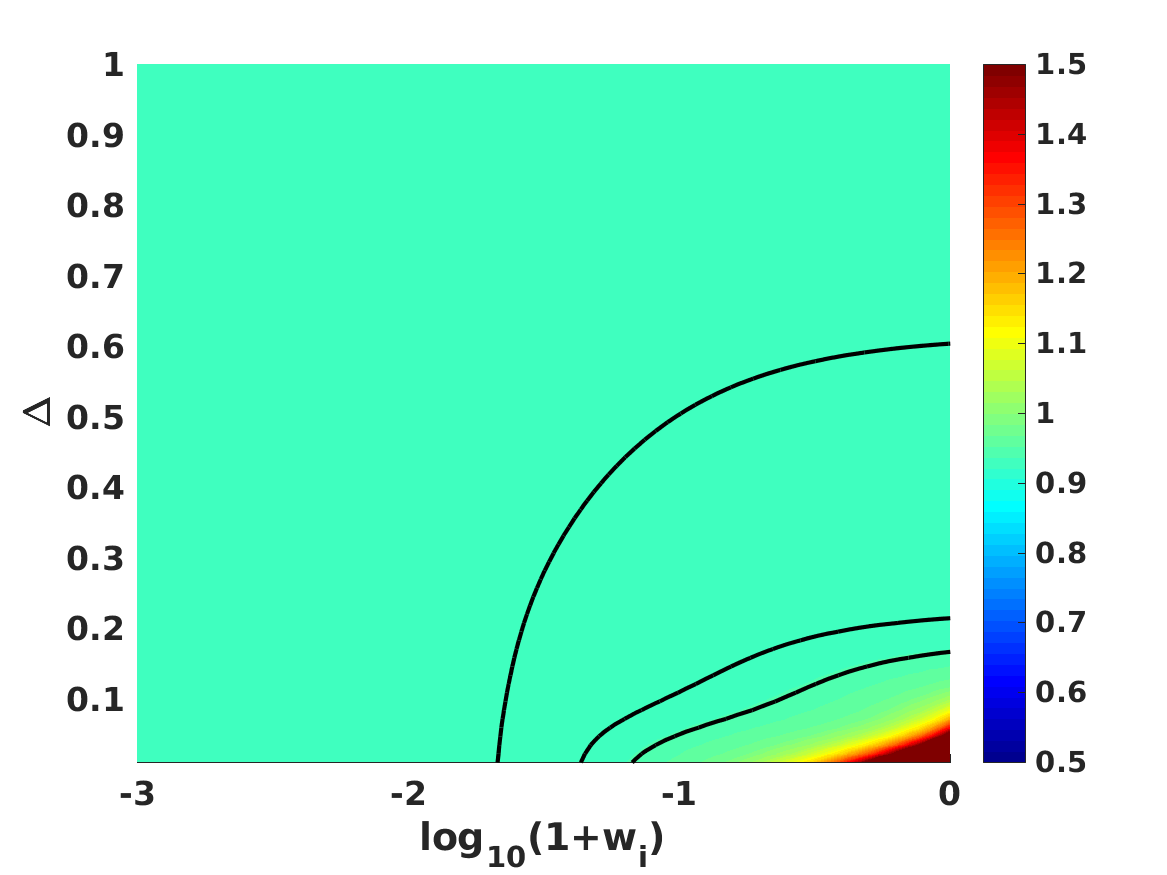}
\caption{Constraints in the ($w_0$,$w_i$) and ($\Delta$,$w_i$) planes (left and right panels, respectively) for the three-parameter Model L. The top panels correspond to a uniform prior on $w_i$, while the bottom ones correspond to a uniform prior on $\log_{10}{(1+w_0)}$. In all cases the black contours denote the one, two and three-sigma confidence levels, while the colourmap depicts the reduced chi-square of the fit for each set of model parameters (dark red corresponds to a reduced chi-square of 1.5 or larger).}
\label{Fig6}%
\end{figure*}

\begin{figure}
\centering
\includegraphics[width=8cm]{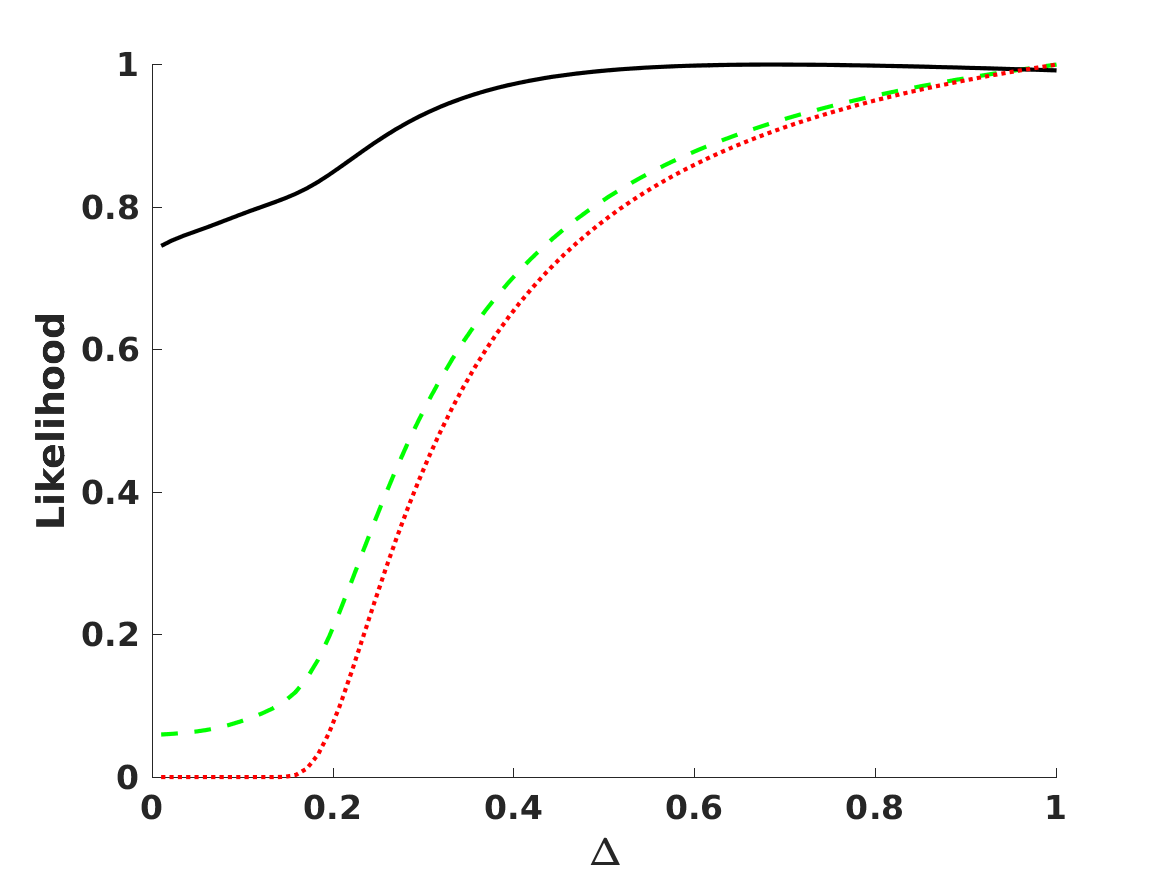}
\includegraphics[width=8cm]{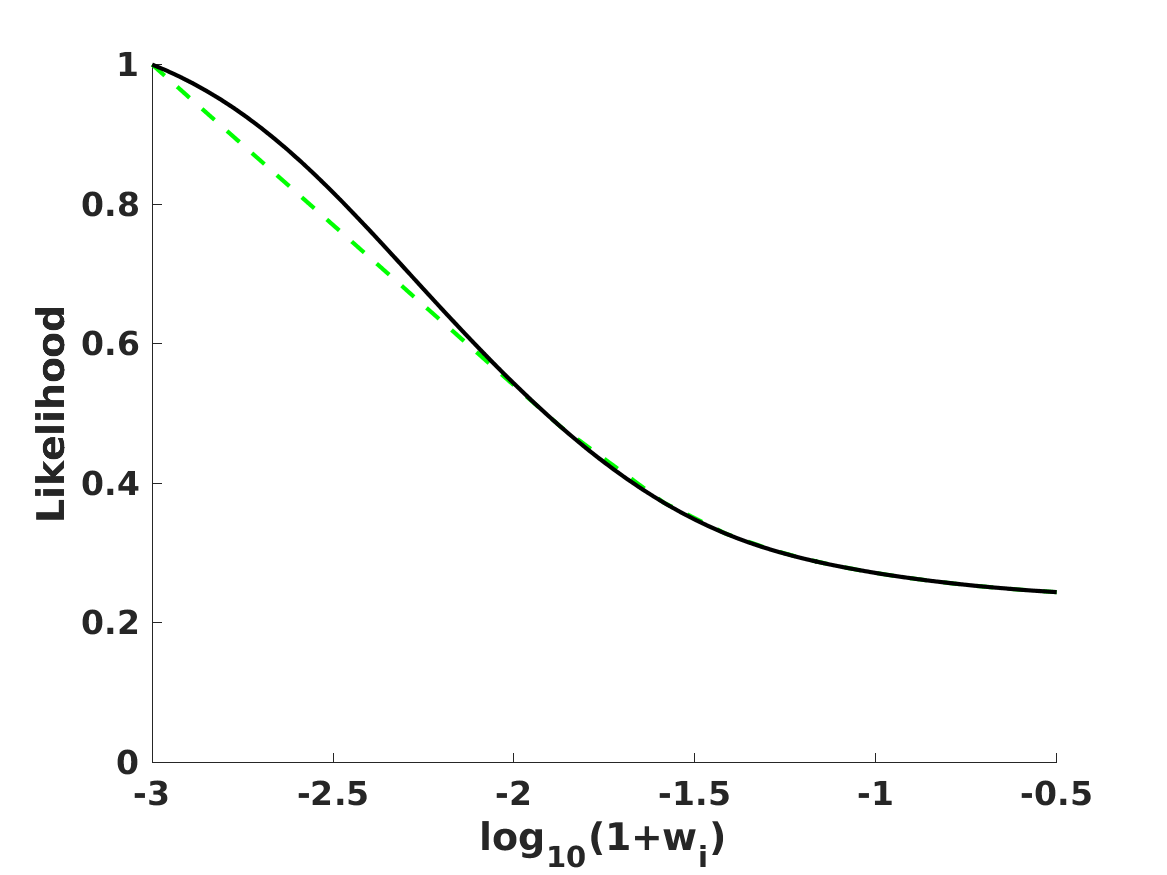}
\includegraphics[width=8cm]{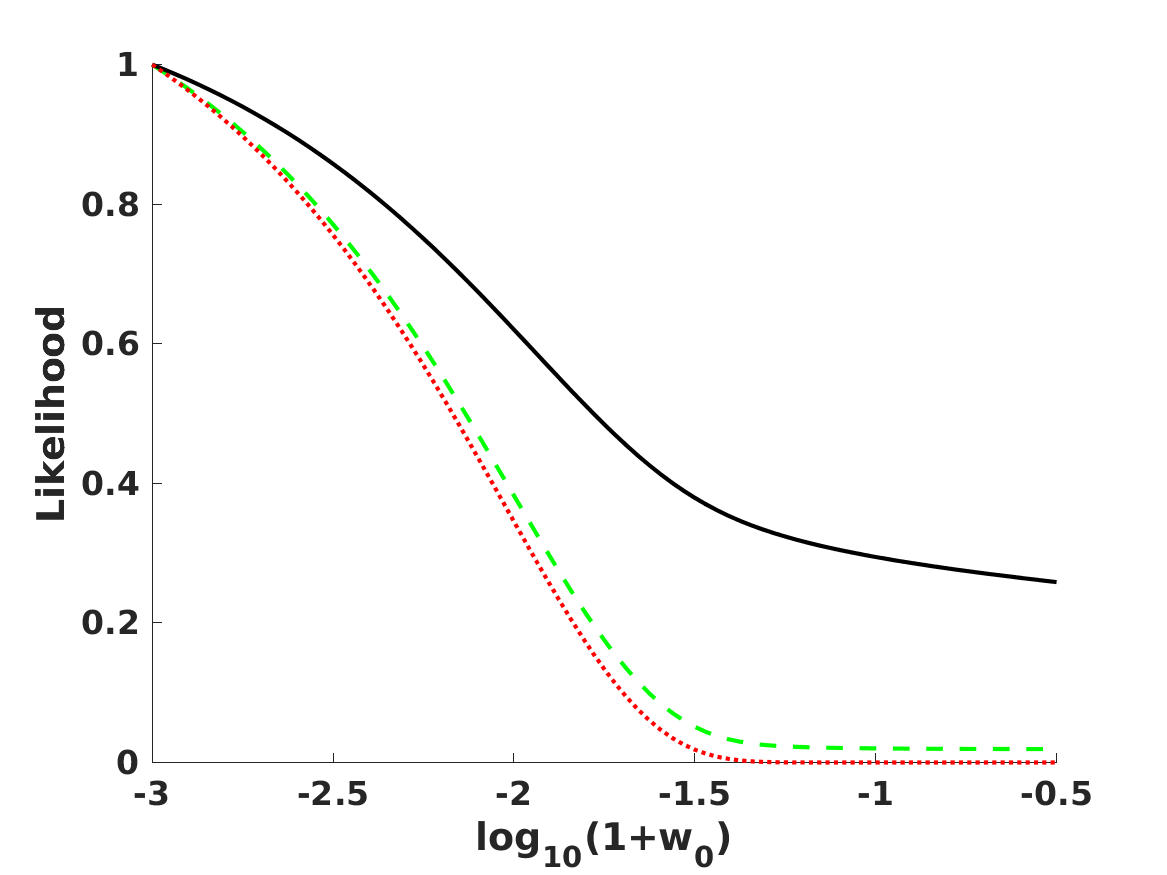}
\caption{Posterior likelihoods for the transition width $\Delta$ and the values of the dark energy equation of state in the asymptotic past, $w_i$ and the present day, $w_0$, with the other parameters marginalised, in the three-parameter version of the L model. The black solid and green dashed lines correspond respectively to the cases of logarithmic and uniform priors on $w_i$, as detailed in the text; for comparison the likelihood for the case of a constant $w_i=0$ is also shown by the red dotted lines. In all cases the Hubble constant has been kept fixed at $H_0=70$ km/s/Mpc.}
\label{Fig7}%
\end{figure}

The results of these analyses are presented in Figures \ref{Fig6} and \ref{Fig7} and Table \ref{table2}. As expected, the extended parameter space weakens the constraints on individual parameters, though this occurs in an interesting way. For the case of a flat (uniform) prior one can still obtain two-sigma bounds on $w_0$ and $\Delta$, though in Table \ref{table2} we report one-sigma bounds since $w_i$ is unconstrained at two-sigma (but still constrained at one-sigma). On the other hand, for a log prior $w_i$ and $w_0$ are still constrained at one sigma (but not two), while $\Delta$ is totally unconstrained.

The physical reason for this behaviour is clear. In the two-parameter case, with a fixed value $w_i=0$ the data strongly prefers a dark energy equation of state at low redshifts (and in particular today, $w_0$) close to that of a cosmological constant, and therefore the putative transition is pushed to high redshifts (into the matter era). In the three parameter case, and despite the correlations between the three parameters, both values of the equation of state ($w_i$ and $w_0$) are constrained to be close to $w_\Lambda=-1$ (though at a lower level of statistical significance), and in such a case there is effectively no transition, so the transition width $\Delta$ is unconstrained. As is clear from Figure \ref{Fig7}, the choice of a linear or logarithmic prior for $w_i$ does affect the posteriors, due to the effects of volume weighting.

The fact that in the case of a logarithmic prior the constraint on $w_i$ is slightly stronger than that of $w_0$ is worthy of comment. A simple but physically meaningful way to classify dynamical dark energy models is to divide them into freezing and thawing ones, depending of the behaviour of the dark energy equation of state $w(z)$ \citep{Caldwell}: qualitatively, in the former ones $w(z)$ is approaching $-1$ today, while in the latter it is moving away from it. Recent observational constraints such as those from the Planck mission in combination with other probes \citep{Ade:2015rim} may be interpreted as slightly preferring thawing models, as pointed out in \citet{LinderFT}. In our case, models with $w_i<w_0$ are thawing models so - with the caveats of the choice of prior and the comparatively low level of statistical significance - our results are also consistent with this interpretation.

\begin{table}
\caption{One-sigma ($68.3\%$ confidence level) constraints on the dark energy equation of state parameters $w_i$, $w_0$ and $\Delta$ (marginalised over the others), for the three-parameter dark energy L model described in the text, with a uniform prior on $w_i$ or on $\log_{10}(1+w_i)$. For comparison the one-sigma constraints on $w_0$ and $\Delta$ for the two-parameter L model with $w_i=0$ discussed in Section 3 are also listed.}
\label{table2}
\centering
\begin{tabular}{| c | c | c | c |}
\hline
Parameter & $w_i=0$ & Flat prior & Log prior  \\
\hline
$\Delta$ & >0.37 & >0.34 & Unconstrained \\
\hline
$\log_{10}(1+w_0)$ & <-2.29 & <-2.26 & <-1.97 \\
\hline
$w_i$ & N/A & <-0.99 & - \\
$\log_{10}(1+w_i)$ & N/A & - & <-2.12 \\
\hline
\end{tabular}
\end{table}


\section{Discussion}

It is enlightening to compare our results with those of previous authors, as a means to ascertain how the various data sets and modelling assumptions influence the resulting constraints. We briefly discuss other relevant works in the context of our results.

The work of \citet{Bassett} provided the initial motivation for our own work. Their analysis has three free parameters, the transition redshift $z_t$, a future value of the dark energy equation of state which is approximately $w_f$ (and indeed, given their other choices of parameters it effectively coincides with $w_0$), and the dark energy density $\Omega_Q$. They fixed $w_i=0$ (as did we in the first part of our article), but further restricted the transition width to obey $\Delta=z_t/30$. Moreover, they assume a flat universe (so $\Omega_m+\Omega_Q=1$) and a Hubble constant $H_0=65$ km/s/Mpc.

Using a combination of CMB, Type Ia supernova and large scale structure data then available, \citet{Bassett} find (at the $68.3\%$ confidence level) $z_t=2.0^{+2.2}_{-0.8}$, $w_f<-0.8$ and $\Omega_Q=0.73^{+0.02}_{-0.04}$. It is clear (and discussed therein) that these constraints mainly come from the CMB: the acoustic peaks and integrated Sachs-Wolfe effect are both affected in these models. The relatively small Type Ia supernova data set then available carries a fairly small statistical weight. In any case, their data sets have a relatively small sensitivity to high redshifts, so to some extent their constraint on $z_t$ can be interpreted as a lower limit, which given their imposed relation between $z_t$ and $\Delta$ would be $\Delta>0.04$, which is comparable to but weaker than our two-parameter result.

A different kind of transition was studied by \citet{Lazkoz}, who considered a class of so-called unified dark energy models, in which dark matter and dark energy are assumed to be different manifestations of the same underlying component which behaves as the former at high redshifts and as the latter at low redshifts. In \citet{Lazkoz} the transition is phenomenologically described by a Heaviside-like function, with the behaviour of the dark energy equation of state before the transition assumed to be $w_i=0$, and the low-redshift behaviour being $w_0=-1$. By using a combination of CMB, galaxy clustering and Type Ia supernova data, they find a best-fit value for the effective redshift of the transition $z_t=4.7\pm0.6$ (at the $68.3\%$ confidence level), though from the point of view of Bayesian evidence there is no clear preference for this model over standard $\Lambda$CDM.

Another four-parameter model for dynamical dark energy was recently proposed by \citet{Jaber1}. This is an extension of the two-parameter CPL model \citep{CPL1,CPL2} with two additional parameters describing the characteristic redshift of the transition between the early and late time behaviours, and the steepness of this transition. Constraints on this model, from a combination of CMB and BAO data together with the local value of the Hubble parameter \citep{Riess:2016jrr} have been recently discussed in \citet{Jaber2}. The four parameters are allowed to vary with stated priors, for example on $z_t=[0,3]$, $w_0=[-1,0]$ and $w_i=[-1,0]$. The authors then find $w_0=-0.96^{+.022}_{-0.17}$ and $w_i=0.00^{+0.04}_{-0.02}$ (despite their statement on the priors on $w_0$ and $w_i$), with a preferred transition redshift $z_t=1.3^{+1.4}_{-0.4}$. Again there is no statistical preference of this model with respect to $\Lambda$CDM. The fact that in their Table 2 they find a very tight constraint $w_i=0.00^{+0.04}_{-0.02}$ from combining CMB, BAO and $H_0$ is intriguing, especially because looking at their Figure 3 the one-sigma confidence level includes $w_i=-1$. The latter is also consistent with their statement that if they just combine BAO with CMB or BAO with $H_0$ they obtain upper bounds consistent with $w_i=-1$.

The full four-parameter model of \citet{Linder:2005ne} has also been studied recently by \cite{Marcondes} (who in addition study two other parametrisations). Here the four parameters are allowed to vary with ample priors (including both canonical and phantom equations of state), and the data used includes CMB, BAO, type Ia supernovas, cosmic chronometers, and the local value of the Hubble parameter. They find tight constraints on the current dark energy equation of state (specifically $w_f=-1.04^{+0.10}_{-0.36}$ at the $68.3\%$ confidence level for the model of \citet{Linder:2005ne}), much weaker constraints on $w_i$ (though consistent with $w_i=-1$) and an effectively unconstrained transition redshift. While their results are consistent with ours, they also confirm the expectation that current data (even including the CMB) is not powerful enough to strongly constrain a four-parameter dark energy model.

More recently \citet{Durrive} have used CMB, Type Ia supernova and BAO data to constrain several classes of canonical quintessence dark energy models. One of the models studied coincides with our Model L since, as first discussed in \citet{Linder:2007wa}, this phenomenological parametrisation is a good approximation to the evolution of the dark energy equation of state in the case of a particular sub-class of quintessence models known as scaling freezing models (for which there is no exact analytic expression for the evolution of $w(z)$). However, they fix the width of the transition to be $\Delta=1/3$, on the grounds that this is the value for which this parametrisation best approximates the dynamics of the said scaling freezing models.

Under these assumptions, \citet{Durrive} obtain a lower bound $z_t>8.1$, at the $95.4\%$ confidence level, and marginalising over all other parameters. No constraints are explicitly given for the other parameters, but it is clear from the text that a cosmological constant is fully consistent with the data. As in the previous case, this constraint is again dominated by the CMB data. While this constraint is stronger than the one obtained in the previous section, our work does show that current low redshift data is already stringent enough to effectively push any hypothetical transition into the matter era. We allow more freedom to the behaviour of the dark energy equation of state (that is, we do not fix $\Delta$), although we are more restrictive in fixing the other cosmological parameters.

Finally, let us again compare our results with those of  \citet{DiValentino}, who have explored and constrained a particular class of vacuum metamorphosis models - first discussed in \citet{Caldwell:2005xb} - where both asymptotic values of the dark energy equation of state correspond to a cosmological constant ($w_i=w_f=-1$, in the notation we have defined above) while during the low-redshift transition the equation of state is in the phantom branch, $w(z)<-1$. For this reason their results are not directly comparable with ours, other than in the general sense that opposite classes of models are being explored. Nevertheless, we note that by using CMB data together with local measurements of the Hubble constant, the authors claim that this class of models can alleviate the $H_0$ tension that seemingly exists for the $\Lambda$CDM model.

In their model the effective transition redshift is not an independent parameter but depends on the matter density and the ratio of a mass-scale to the Hubble constant. Because of this, \citet{DiValentino} do not explicitly provide constraints on the transition redshift, but it is curious to note that for their best-fit model the transition redshift would be $z_t\sim1.2$, which is again in the matter era and comparable to our constraints.


\section{Conclusions}

We have studied classes of phenomenological but physically motivated two and three parameter models where the dark energy equation of state undergoes a (possibly) rapid transition at low redshifts. Such transitions might be associated with the onset of the acceleration phase itself. By using  Type Ia supernova and Hubble parameter measurements, we have provided updated constraints on the redshift and steepness of these possible transitions as well as on the values of the dark energy equation of state at the present day and in the asymptotic past in these models.

Our results confirm and update those of previous works, showing that such transitions are tightly constrained. The present-day value of the dark energy equation of state, $w_0$ is always constrained to be close to that of a cosmological constant, $w_\Lambda=-1$. In the two-parameter case where $w_i$ is fixed at zero, the low-redshift data pushes the transition to high redhsifts. Effectively, this transition must happen within the matter era. Nevertheless, a value of $w_i=0$ is also constrained by other data, and especially by the CMB. Relaxing this assumption and allowing $ w_i$ to be a third free model parameter, then it is also constrained to be close to $w_\Lambda=-1$ so there is effectively no transition. Our constraints have a mild dependence of the specific parametrisation being used and on the choice of priors for the model parameters, but should be relatively insensitive to the choice of the Hubble parameter $H_0$.

We conclude that the dark energy equation of state near the present day must be very similar to that of a cosmological constant, and any significant deviations from this behaviour can only occur in the deep matter era. In particular, at the level of the behaviour of the dark energy equation of state there is effectively no room for a phase transition associated with the onset of the acceleration phase. Therefore, if one wants to identify possible deviations from a cosmological constant, it is important to develop observational tools that are capable of probing this behaviour in the deep matter era. Two such promising examples are tests of the stability of fundamental couplings \citep{Martins:2017yxk} and direct measurements of the expansion of the universe \citep{Liske:2008ph,Martins:2016bbi}. We leave discussion of the potential of these observables to further constrain these models to subsequent work.

\begin{acknowledgements}
We are grateful to Ana Catarina Leite, Ferran Delg\`a, Sara Raposeiro, and Sofia Cardoso for helpful discussions on the subject of this work. This work was financed by FEDER---Fundo Europeu de Desenvolvimento Regional funds through the COMPETE 2020---Operacional Programme for Competitiveness and Internationalisation (POCI), and by Portuguese funds through FCT---Funda\c c\~ao para a Ci\^encia e a Tecnologia in the framework of the project POCI-01-0145-FEDER-028987. C.J.M. is supported by an FCT Research Professorship, contract reference IF/00064/2012, funded by FCT/MCTES (Portugal) and POPH/FSE (EC). M.P.C. acknowledges financial support from Programa Joves i Ci\`encia, funded by Fundaci\'o Catalunya-La Pedrera (Spain).
\end{acknowledgements}

\bibliographystyle{aa} 
\bibliography{dark} 

\end{document}